\documentclass[namedreferences]{SolarPhysics}
\usepackage[optionalrh,solaenum,natbib]{spr-sola-addons} 
\usepackage{graphicx}                    
\usepackage{color}                       
\usepackage{url,bm}

\graphicspath{{./fig/}{./png/}}
\newcommand{\EQ}{\begin{equation}}
\newcommand{\EN}{\end{equation}}
\newcommand{\EQA}{\begin{eqnarray}}
\newcommand{\ENA}{\end{eqnarray}}

\newcommand{\Eq}[1]{Equation~(\ref{#1})}

\newcommand{\App}[1]{Appendix~\ref{#1}}

\newcommand{\Fig}[1]{Figure~\ref{#1}}
\newcommand{\FFig}[1]{Figure~\ref{#1}}
\newcommand{\Figs}[2]{Figures~\ref{#1} and \ref{#2}}

\newcommand{\bra}[1]{\langle #1\rangle}

\newcommand{\meanrho}{\overline{\rho}}

{}
{}
\newcommand{\meanFFFF}{\overline{\mbox{\boldmath ${\cal F}$}}{}}{}

\newcommand{\meanSSSS}{\overline{\mbox{\boldmath ${\mathsf S}$}} {}}

\newcommand{\meanSSS}{\overline{\mathsf{S}}}
{}
{}
{}
{}
{}
{}
{}
{}
\newcommand{\meanAA}{\overline{\mbox{\boldmath $A$}}{}}{}
\newcommand{\meanBB}{\overline{\mbox{\boldmath $B$}}{}}{}
{}
{}
{}
{}
{}
{}
{}
{}
\newcommand{\meanJJ}{\overline{\mbox{\boldmath $J$}}{}}{}
{}
\newcommand{\meanUU}{\overline{\bm{U}}}

{}
{}
{}

\newcommand{\meanB}{\overline{B}}

\newcommand{\meanU}{\overline{U}}

\newcommand{\meanp}{\overline{p}}

{}

{}
{}

\newcommand{\uu}{\mbox{\boldmath $u$} {}}
\newcommand{\UU}{\mbox{\boldmath $U$} {}}

\def\bb{\bm{b}}
\newcommand{\BB}{\mbox{\boldmath $B$} {}}

\newcommand{\JJ}{\mbox{\boldmath $J$} {}}

\newcommand{\AAA}{\mbox{\boldmath $A$} {}}

\newcommand{\ff}{\mbox{\boldmath $f$} {}}

\newcommand{\grav}{\mbox{\boldmath $g$} {}}
\newcommand{\nab}{\mbox{\boldmath $\nabla$} {}}

\newcommand{\SSSS}{\mbox{\boldmath ${\sf S}$} {}}

\newcommand{\DD}{{\rm D} {}}

\newcommand{\dd}{{\rm d} {}}

\newcommand{\const}{{\rm const}  {}}

\def\Pm{\mbox{\rm Pr}_M}
\def\Rm{\mbox{\rm Re}_M}

\def\Rey{\mbox{\rm Re}}

\def\cs{c_{\rm s}}
\def\qpz{q_{\rm p0}}
\def\qp{q_{\rm p}}
\def\betap{\beta_{\rm p}}

\def\betastar{\beta_{\star}}

\def\Peff{{\cal P}_{\rm eff}}
\def\Pmin{{\cal P}_{\rm min}}
\def\qs{q_{\rm s}}

\def\vA{v_{\rm A}}

\def\kf{k_{\rm f}}

\def\urms{u_{\rm rms}}

\def\nut{\nu_{\rm t}}

\def\etat{\eta_{\it t}}

\def\Beq{B_{\rm eq}}
\def\Beqz{B_{\rm eq0}}

\def\half{{\textstyle{1\over2}}}

\def\onethird{{\textstyle{1\over3}}}

\newcommand{\Mm}{\,{\rm Mm}}

\newcommand{\yapj}[3]{ #1, {ApJ,} {#2}, #3}

\newcommand{\yapjl}[3]{ #1, {ApJ,} {#2}, #3}

\newcommand{\yan}[3]{ #1, {Astron.\ Nachr.,} {#2}, #3}

\newcommand{\yana}[3]{ #1, {A\&A,} {#2}, #3}

\newcommand{\yjfm}[3]{ #1, {J.\ Fluid Mech.,} {#2}, #3}

\newcommand{\ysovl}[3]{ #1, {Sov.\ Astron.\ Lett.,} {#2}, #3}

\newcommand{\yjetp}[3]{ #1, {Sov.\ Phys.\ JETP,} {#2}, #3}

\newcommand{\yanf}[3]{ #1, {Ann. Rev. Fluid Mech.,} {#2}, #3}

\newcommand{\ymn}[3]{ #1, {MNRAS,} {#2}, #3}
\newcommand{\ynat}[3]{ #1, {Nature,} {#2}, #3}

\newcommand{\ysci}[3]{ #1, {Science,} {#2}, #3}
\newcommand{\ysph}[3]{ #1, {Solar Phys.,} {#2}, #3}

\newcommand{\ypre}[3]{ #1, {Phys.\ Rev.\ E,} {#2}, #3}

\newcommand{\ybook}[3]{ #1, {#2} (#3)}

\newcommand{\pmn}[1]{ #1, {MNRAS}, to be published}
\newcommand{\psph}[1]{ #1, {Sol.\ Phys.}, to be published}

\newcommand{\papj}[1]{ #1, {ApJ}, to be published}

\begin{document}

\begin{article}

\begin{opening}

\title{Active region formation through
the negative effective magnetic pressure instability}

%
\author{Koen Kemel$^{1,2}$\sep
Axel Brandenburg$^{1,2}$\sep
Nathan Kleeorin$^{3,1}$\sep
Dhrubaditya Mitra$^1$\sep
Igor Rogachevskii$^{3,1}$
}
%
\runningauthor{Kemel {\it et al.}}
\runningtitle{Negative effective magnetic pressure instability}
%
  \institute{$^{1}$ Nordita, AlbaNova University Center,
    Roslagstullsbacken 23, SE-10691 Stockholm, Sweden,
    email: brandenb@nordita.org \\
   $^{2}$ Department of Astronomy, Stockholm University, SE-10691
   Stockholm, Sweden\\
   $^{3}$ Department of Mechanical Engineering, Ben-Gurion University of the Negev, POB 653,
   Beer-Sheva 84105, Israel\\
 }
\begin{abstract}
The negative effective magnetic pressure instability operates on scales
encompassing many turbulent eddies and is here discussed in connection
with the formation of active regions near the surface layers of the Sun.
This instability is related to the negative contribution of turbulence to
the mean magnetic pressure that causes the formation of large-scale
magnetic structures.
For an isothermal layer, direct numerical simulations and mean-field
simulations of this phenomenon
are shown to agree in many details in that their onset
occurs at the same depth. This depth increases with increasing field strength,
such that the maximum growth rate of this instability
is independent of the field strength, provided
the magnetic structures are fully contained within the domain.
A linear stability analysis is shown to support this finding.
The instability also leads to a redistribution of turbulent intensity
and gas pressure that could provide direct observational signatures.
\end{abstract}
\end{opening}

\section{Introduction}

Active region formation in the Sun is traditionally thought to be a
deeply rooted phenomenon, because their size ($\sim100\Mm$) is much
larger than the naturally occurring scales in the surface layers of
the convection zone ($\sim1$--$10\Mm$); see \cite{Golub}.
They are also long-lived
($\sim1/2$ year), which seems unnaturally long if associated with
the near-surface layers ($40\Mm$ depth) where typical time scales
are about a day. On the other hand, a deeply rooted formation
scenario for active regions has the problem that the azimuthal pattern
speed of active regions does not match the
angular velocity at great depth.
Other difficulties concern the strong field strength inferred for the
tachocline to explain the observed tilt angles and the fact that
magnetic structures expand tremendously during their ascent.
These and several other arguments have led to the consideration of solar
activity as a shallow phenomenon; see \cite{B05} for details.
As a possible mechanism for producing magnetic flux concentrations of the
form of active regions, the negative effective magnetic pressure instability
has been discussed \citep{KRR89,KRR90,KR94,KMR96,RK07,BKR10,BKKMR}.
Of course a magnetic field $\BB$ always gives rise to a positive
magnetic pressure, $\BB^2/2\mu_0$,
where $\mu_0$ is the vacuum permeability.  In a turbulent medium,
however, magnetic fields also suppress the turbulence and thus decrease
the turbulent pressure $\rho\uu^2/3$, and modify the pressure caused
by magnetic fluctuations $\bb^2/6\mu_0$.
Here, $\uu$ and $\bb$ are velocity and magnetic fluctuations,
$\rho$ is the density, $\mu_0$ is the vacuum permeability,
and the coefficients in the turbulent fluid and magnetic pressure are
given for isotropic turbulence.
Magnetic fluctuations can be due to both small-scale dynamo action
as well as tangling of a large-scale field, $\meanBB$.
The total field is thus $\BB=\meanBB+\bb$.
The sum of both effects,
$p_{\rm turb}=\overline{\rho\uu^2}/3+\bra{\bb^2}/6\mu_0$,
is positive definite, but it depends on $\meanBB$, and $p_{\rm turb}$
tends to decline as $\meanB\equiv|\meanBB|$ increases.
Indeed, $p_{\rm turb}=2E_K/3-\bra{\bb^2}/6\mu_0$, where $E_K=\overline{\rho\uu^2}/2
+\bra{\bb^2}/2\mu_0 \approx$ const, so that the change of the turbulent
pressure is negative $(\delta p_{\rm turb}<0)$ when the magnetic fluctuations
are generated by tangling of the mean magnetic field by the velocity fluctuations
at the expanse of turbulent kinetic energy \citep{KRR90,BKKMR}.
Thus, we write
\EQ
p_{\rm turb}(\meanB)=p_{\rm turb}(0)-\qp(\meanB)\meanB^2/2\mu_0,
\EN
where $p_{\rm turb}(0)$ is the turbulent pressure at zero mean field.
The pressure $p_{\rm turb}(0)$
only includes those contributions from $\bb^2$ that are associated with
small-scale dynamo action, but not those resulting from the mean field.
The relevant magnetic pressure in the evolution equation for the mean flow,
$\meanUU$, is then not just $\meanB^2/2\mu_0$, but it is
affected by the $\meanB$ dependence of $p_{\rm turb}$, i.e.,
it depends on
\EQ
p_{\rm turb}(\meanB)+\meanB^2/2\mu_0=
p_{\rm turb}(0)+[1-\qp(\meanB)]\meanB^2/2\mu_0,
\EN
which is also still positive, but $1-\qp(\meanB)$
may well become negative, which is what we call a
{\em negative effective magnetic pressure}.
Consequently, the expression
\EQ
p_{\rm eff}=(1-\qp)\meanBB^2/2\mu_0
\EN
is referred to as the effective magnetic pressure.
In addition, there is also the gas pressure $p_{\rm gas}$.
Once the effective magnetic pressure drives a mean flow, the
gas density changes, and as a consequence the gas pressure,
so as to re-establish approximate total pressure balance.
Therefore, $p_{\rm gas}$ and $\rho$ will also depend on $\meanB$.

In the presence of gravity, the properties of magnetic buoyancy
are drastically altered by a negative effective magnetic pressure.
In the following we illustrate how this can lead to an instability.
Since the flow velocities are highly subsonic, we can make
the anelastic approximation for low Mach numbers,
i.e., $\nab\cdot\meanrho\meanUU=0$.
This leads to $\nab\cdot\meanUU+\meanUU\cdot\nab\ln\meanrho=0$, or
\EQ
\nab\cdot\meanUU={\meanU_z\over H_\rho},
\label{divu}
\EN
where we have used the density scale height $H_\rho$,
so that $\nab\ln\meanrho=(0,0,-1/H_\rho)$.
This equation shows that a downward motion $\meanU_z<0$ leads to
a compression, $\nab\cdot\meanUU<0$.
This enhances an applied field locally.
We consider an applied equilibrium magnetic field of the form $(0,B_0,0)$
and the mean field has only a $y$ component,
i.e., $\meanBB=(0,\meanB_y(x,z),0)$, so we have
\EQ
{\DD\meanB_y\over\DD t}=-\meanB_y\nab\cdot\meanUU,
\label{DByDt}
\EN
where $\DD/\DD t=\partial/\partial+\meanUU\cdot\nab$
is the advective derivative.
Note that for a magnetic field with only a $y$ component, but
$\partial/\partial y=0$, there is no stretching term, so
there is no term of the form $\meanBB\cdot\nab\meanUU$.
Using \Eq{divu}, and linearizing \Eq{DByDt} around $\meanUU=\bm{0}$
and $\meanBB=\BB_0$, we have
\EQ
{\partial\meanB_{1y}\over\partial t}=-B_0{\meanU_{1z}\over H_\rho},
\label{induct-eq}
\EN
where subscripts 1 denote linearized quantities.
The vertical velocity perturbation $\meanU_{1z}$ is caused by
magnetic buoyancy.
Assuming total pressure equilibrium, $p_{\rm gas}+p_{\rm eff}=\const$,
we see that an increase in the effective magnetic pressure causes
a {\it decrease} is the gas pressure, i.e.,
$\delta p_{\rm gas}=-\delta p_{\rm eff}$, just like in the regular
magnetic buoyancy instability.
Therefore, the Archimedian buoyancy force is
\EQ
-{\delta\rho\over\rho}g=-{\delta p_{\rm gas}\over p_{\rm gas}}g
={\delta p_{\rm eff}\over\rho\cs^2}g
={\dd p_{\rm eff}\over\dd\meanB^2}{\delta\meanB^2\over\rho\cs^2}g,
\EN
where we have used $p_{\rm gas}=\rho\cs^2$ for an isothermal gas.
In the regular magnetic buoyancy instability, without turbulence effects,
we have $2\mu_0\dd p_{\rm eff}/\dd{\meanB}{}^2=1$.
In the domain where the negative effective magnetic pressure effect causes
$\dd p_{\rm eff}/\dd{\meanB}{}^2$
to be negative, a magnetic field enhancement leads to
a further reduction of the local pressure, which is compensated by horizontal
inflows, increasing density (and field strength),
making this fluid parcel heavier, causing it to sink.
Inversely, a local field reduction causes outflows and rises until
it reaches the region where this feedback reverses.
Thus, the instability loop is closed by considering the momentum equation
in its linearized form
\EQ
{\partial\meanU_{1z}\over\partial t}={\dd p_{\rm eff}\over\dd\meanB^2}\,
{2B_0\meanB_{1y}\over \rho\cs^2} \,g.
\EN
Using $\cs^2/g=H_\rho$ for an isothermal atmosphere, we then find
the dispersion relation for the growth rate $\lambda$
of the resulting instability
\EQ
\lambda={\vA\over H_\rho}\sqrt{-2 \mu_0\, p_{\rm eff}'}-\etat k^2,
\label{disper}
\EN
where $\vA=B_0/\sqrt{\mu_0\rho}$ is the Alfv\'en speed and
\EQ
2\mu_0 \, p_{\rm eff}'=2\mu_0 \, \dd p_{\rm eff}/\dd{\meanB}^2=1-\qp-\dd\qp/\dd\ln\meanB^2
\EN
is twice the derivative of the effective magnetic pressure.
We have here also included the effects of turbulent magnetic diffusivity
$\etat$ and turbulent magnetic viscosity $\nut$, assuming $\nut/\etat=1$.
Here, $k$ is the effective wavenumber.
A proper derivation of the growth rate of the instability,
but again without including turbulent magnetic diffusivity
and turbulent magnetic viscosity, is given in \App{LinTheo}.

The negative contribution of turbulence to the mean magnetic pressure
and the resulting large-scale instability has been predicted long
ago \citep{KRR90,KMR96,KR94}. However, this instability
has been detected in DNS only recently \citep{BKKMR,KBKMR12}.
This large-scale instability is called the negative effective magnetic
pressure instability, or NEMPI, for short.

\Eq{disper} demonstrates that stronger stratification and thus a
smaller scale height leads to an increased growth rate
of the instability.
This was qualitatively confirmed by \cite{KBKR12}.
Using numerical solutions of the full mean-field equations,
they found furthermore that the
maximum growth rate of the instability
is actually independent of $\vA$.
This seems to be at odds with the \Eq{disper}.
To understand this, we use the following fit formula for $\qp$:
\EQ
\qp(\beta)={\betastar^2\over\betap^2+\beta^2},
\EN
where $\beta=\meanB/\Beq$ and $\Beq=\sqrt{\mu_0\rho}\urms$
is the equipartition field strength.
Thus, for $\betastar\gg\beta\gg\betap$, we have
\EQ
\lambda\approx\betastar{\urms\over H_\rho}-\etat k^2,
\label{E1}
\EN
so the growth rate is indeed independent of the imposed field strength.

In a mean-field model, $\urms$ is normally expressed in terms of
$\etat=\urms/3\kf$, so \Eq{E1} turns into
\EQ
{\lambda\over\etat k^2}\approx3\betastar{\kf/k\over k H_\rho}-1,
\EN
which illustrates immediately the importance of large enough
scale separation, i.e., large enough values of $\kf/k$.

The purpose of this paper is to show that
NEMPI can work over a range of different field strength.
Such a result was recently predicted using
the mean-field simulations (MFS) by \cite{KBKR12}.
We shall also investigate the close connection between mean
field and the resulting effective magnetic pressure in the plane
perpendicular to the mean field.
Here we focus on a series of simulations with different field strengths,
but for a fixed value of the magnetic Reynolds number and fixed value
of the scale separation ratio.
For a study of the dependence on magnetic Reynolds number and
on scale separation ratio, but fixed field strength, we refer
to the recent work of \cite{KBKMR12}.
In the following, we discuss first direct numerical simulations (DNS)
of NEMPI and turn then to proper mean-field simulations (MFS).
We begin with a simplistic illustration of the nature of NEMPI.

\section{Vertical profile of effective magnetic pressure}

The first successful DNS of NEMPI have been possible under the
assumption of an isothermally stratified layer with an isothermal
equation of state \citep{BKKMR}.
Much of the same physics is also possible in adiabatically stratified
layers, but NEMPI was found in this case only
in mean-field models \citep{BKR10,KBKMR11}.
The isothermal case has conceptual advantages that help us understanding
better the underlying physics of this instability.
We make use of this advantage in the present paper, too.

\begin{figure}[t!]\begin{center}
\includegraphics[width=\columnwidth]{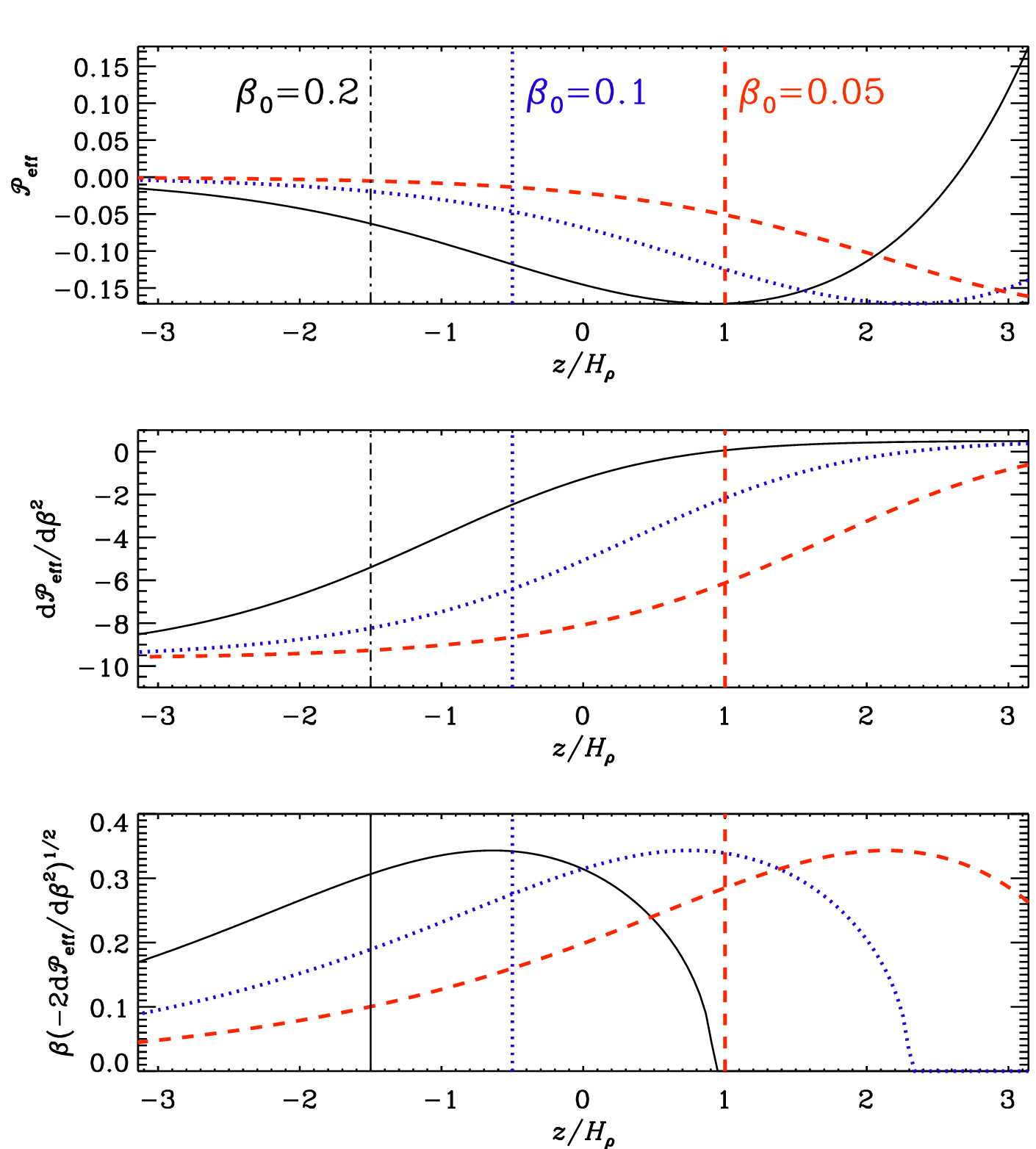}
\end{center}\caption[]{
Profiles of $\Peff$, $\dd\Peff/\dd\beta^2$, and
$\beta(-2\dd\Peff/\dd\beta^2)^{1/2}$
for fit parameters $\qpz=20$ and $\betap=0.167$, and the three
field strengths $B_0/\Beqz=0.05$, 0.1, and 0.2 within the $z$ range
from $-\pi$ to $\pi$, which is consistent with some of the models
discussed below.
The vertical lines of similar line types give the location
where the unstable eigenmodes reaches its peak.
}\label{MF_init}\end{figure}

In most of the isothermal setups studied so far, the rms velocity
is only weakly dependent on
height, so the $z$ variation of $\Beq$ was only caused by that of
$\rho=\rho_0\exp(-z/H_\rho)$.
This allows us then to plot the effective magnetic pressure.
In the following, we introduce the quantity
\EQ
\Peff(\beta)=\half\left[1-\qp(\beta)\right]\beta^2,
\EN
which is the effective magnetic pressure normalized by the
local equipartition field strength $\Beq$,
i.e., $\Peff=p_{\rm eff}/\Beq^2$.
Since $\beta=\beta(z)=\beta_0\exp(z/2H_\rho)$ increases with $z$,
$\Peff(z)$ is small at large depths, reaches a negative minimum
at some depth, and then becomes positive and equal to $\beta^2$.
In \Fig{MF_init} we show vertical profiles of $\Peff$,
$\dd\Peff/\dd\beta^2$, and $\beta(-2\dd\Peff/\dd\beta^2)^{1/2}$
for the fit parameters $\qpz=20$ and $\betap=0.167$ derived later
in this paper,
and the three field strengths $\beta_0\equiv B_0/\Beqz=0.05$, 0.1, and 0.2
within the $z$ range from $-\pi$ to $\pi$, which is also consistent
with the DNS and some of the MFS discussed below.
Here, $\Beqz=\Beq(0)$.

Notice first of all that all three curves of $\Peff$ have minima with
left flanks (negative slopes) within the domain.
As the imposed field is increased, these curves shift downward
(smaller values of $z$).
Thus, we should expect the peak of the instable eigenmode
to appear somewhere along the left flanks of these curves and
that these peaks move further down as the imposed field
is increased.
This is qualitatively reproduced by the DNS and MFS discussed
below, except that the location is consistently a certain
distance below the position where the left flanks have their
steepest gradient.
On the other hand, as is evident from the middle panel of
\Fig{MF_init}, the largest value of $\dd\Peff/\dd\beta^2$
is always achieved at the bottom of the domain.
However, the growth rate of NEMPI
has still a factor proportional to
$\vA=\urms\beta$ in front of it; see \Eq{disper}.
This then confines the instability to a narrow strip within the domain.
In the third panel of \Fig{MF_init} we plot therefore also
$\beta(-2\dd\Peff/\dd\beta^2)^{1/2}$, and their extrema are
now only slightly above the location where DNS and MFS
show a peak in the eigenfunction.
The reason for the remaining discrepancy is not well understood at present.

\section{Onset and saturation of NEMPI in DNS}
\label{DNS}

\subsection{Isothermal setup in DNS}
\label{model}

Following the earlier work of \cite{BKKMR} and \cite{KBKMR12},
we solve the equations for the velocity $\UU$,
the magnetic vector potential $\AAA$, and the density $\rho$,
\begin{equation}
\rho{\DD\UU\over\DD t}=-\cs^2\nab\rho+\JJ\times\BB+\rho(\ff+\grav)
+\nab\cdot(2\nu\rho\SSSS),
\end{equation}
\begin{equation}
{\partial\AAA\over\partial t}=\UU\times\BB+\eta\nabla^2\AAA,
\end{equation}
\begin{equation}
{\partial\rho\over\partial t}=-\nab\cdot\rho\UU,
\end{equation}
where $\nu$ is the kinematic viscosity, $\eta$ is the magnetic diffusivity
due to Spitzer conductivity of the plasma,
$\BB=\BB_0+\nab\times\AAA$ is the magnetic field,
$\BB_0=(0,B_0,0)$ is the imposed uniform field,
$\JJ=\nab\times\BB/\mu_0$ is the current density,
$\mu_0$ is the vacuum permeability,
${\sf S}_{ij}=\half(U_{i,j}+U_{j,i})-\onethird\delta_{ij}\nab\cdot\UU$
is the traceless rate of strain tensor, and commas denote
partial differentiation.
The forcing function $\ff$ consists of random, white-in-time,
plane non-polarized waves with a certain average wavenumber.
The turbulent rms velocity is approximately
independent of $z$ with $\urms=\bra{\uu^2}^{1/2}\approx0.1\,\cs$.
The gravitational acceleration $\grav=(0,0,-g)$ is chosen such that
$k_1 H_\rho=1$, so the density contrast between
bottom and top is $\exp(2\pi)\approx535$.
Here, $H_\rho=\cs^2/g$ is the density scale height.
We consider a domain of size $L_x\times L_y\times L_z$ in
Cartesian coordinates $(x,y,z)$, with periodic boundary conditions in
the $x$ and $y$ directions and stress-free perfectly conducting
boundaries at top and bottom ($z=\pm L_z/2$).
In all cases, we use a scale separation ratio $\kf/k_1$ of 30,
a fluid Reynolds number $\Rey\equiv\urms/\nu\kf$ of 18,
and a magnetic Prandtl number $\Pm=\nu/\eta $ of 0.5.
In our units, $\mu_0=1$ and $\cs=1$.
The value of $B_0$ is specified in units of the volume
averaged value, $\Beqz=\sqrt{\mu_0\rho_0} \, \urms$,
where $\rho_0=\bra{\rho}$ is the volume-averaged density,
which is constant in time.
In addition to visualizations of the actual magnetic field,
we also monitor $\meanB_y$, which is an
average over $y$ and a certain time interval $\Delta t$.
Since the simulations are periodic in the $x$ and $y$ directions,
we sometimes shift the images such that the peak field strength
of NEMPI appears in the middle of the frame.

The simulations are performed with the {\sc Pencil Code},%
\footnote{{\tt http://pencil-code.googlecode.com}}
which uses sixth-order explicit finite differences in space and a
third-order accurate time stepping method.
We use a numerical resolution of $256^3$ mesh points.

\subsection{Results}
\label{DNSResults}

\begin{figure}[t!]\begin{center}
\includegraphics[width=\columnwidth]{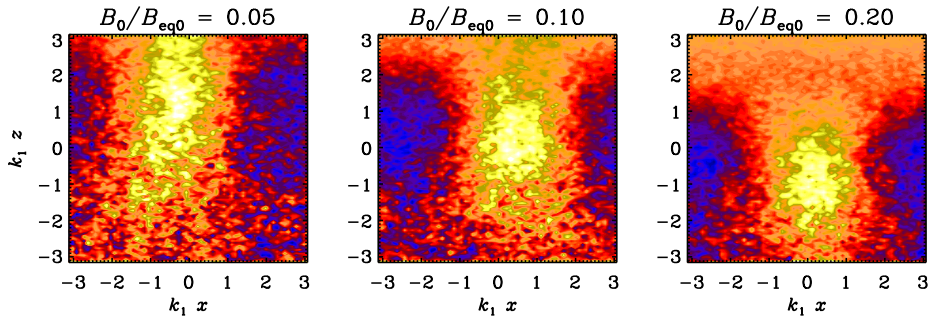}
\end{center}\caption[]{
$\meanB_y$ from  DNS for three values of the imposed field strength at
the end of the linear growth phase of NEMPI for $\Rm=18$ and $\Pm=0.5$.
}\label{DNS_B}\end{figure}

\begin{figure}[t!]\begin{center}
\includegraphics[width=\columnwidth]{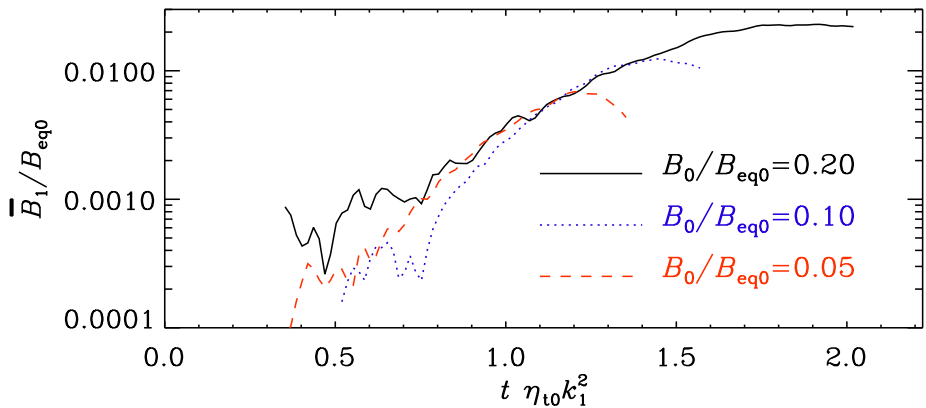}
\end{center}\caption[]{
Growth of the large-scale field strength $\meanB_1$ at the center of
the magnetic structure for three field strengths.
}\label{DNS_grr}\end{figure}

In \Fig{DNS_B} we demonstrate that NEMPI can work over
a range of field strengths.
As we increase the strength of the imposed field, NEMPI develops
at progressively greater depth.
This result was recently obtained for MFS, but is now for the first
time demonstrated in DNS.
\FFig{DNS_grr} shows that the growth of the large-scale field $\meanB_1$
of the magnetic structure is similar for three different field strengths.
Here, $\meanB_1$ has been determined by taking the maximum value of
the mean field in the neighborhood of the position where the flux
concentration later develops.
Note that there is a range over which $\meanB_1$ grows approximately
exponentially, independent of the value of $B_0$.

\begin{figure}[t!]\begin{center}
\includegraphics[width=.8\columnwidth]{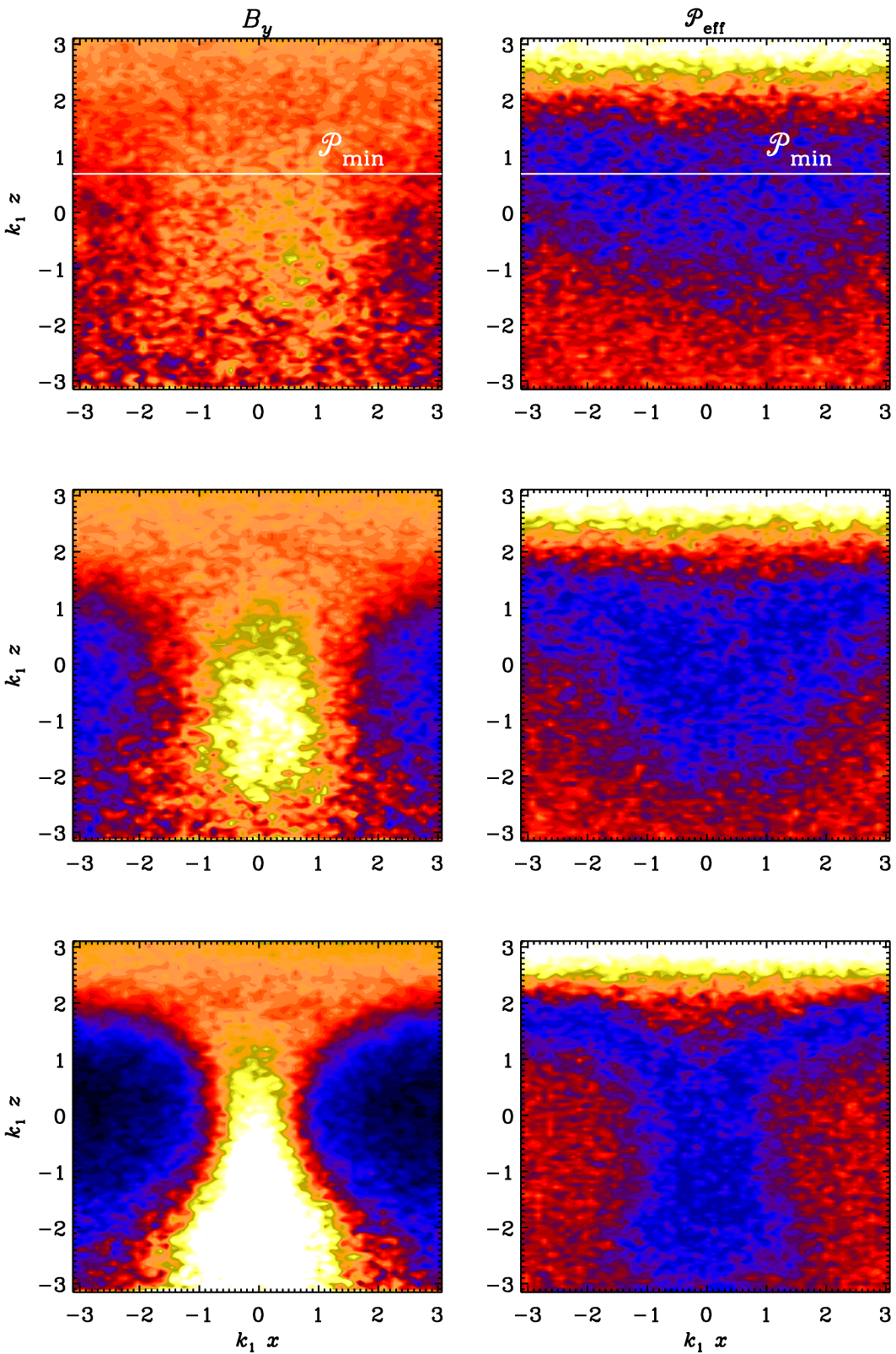}
\end{center}\caption[]{
$\meanB_y$ and $\Peff$ from DNS at three times showing the descent
of the potato sack feature for $\Rm=18$ and $\Pm=0.5$.
}\label{DNS_P}\end{figure}

In \Fig{DNS_P} we show $\meanB_y$ at early, intermediate, and late stages
of the saturation process (left), and compare with visualizations of
$\Peff$ at the same times.
Here, $\Peff=\half(1-\qp)\beta^2$, where $\qp(\beta)$
with $\beta=\meanB/\Beq$ is evaluated from
\EQ
\qp=-2\Delta\overline\Pi_{xx}^{\rm f}/\meanBB^2,
\EN
for $\BB_0=(0,B_0,0)$, and
\EQ
\Delta\overline\Pi_{ii}^{\rm f}
=\meanrho\,(\overline{u_i^2}-\overline{u_{0i}^2})
+\half(\overline{\bb^2}-\overline{\bb_0^2})
-(\overline{b_i^2}-\overline{b_{0i}^2}),
\label{Pi_ii}
\EN
is applied to the $xx$ component of the total stress from the fluctuating
velocity and magnetic fields.
In \Eq{Pi_ii} no summation over the repeated index $i$ is assumed.

In \Fig{DNS_P}, blue shades correspond to low values of $\Peff$ and occur
around the minimum line (marked in white) where $\Peff=\Pmin$.
As time progresses, low values of $\Peff$ are also found at
greater depth as the magnetic flux concentration descends.
The fact that there is a clear spatial correlation between $\meanB_y$ and
$\Peff$ provides strong evidence that the interpretation of the formation
of structures in the stratified turbulence simulations in terms of
NEMPI is indeed the correct one.

The descending structures have previously been referred to as potato
sack structures \citep{BKKMR}, because of their widening cross-section
with greater depth.
When such structures were first seen in MFS \citep{BKR10}, they were
originally thought to be artifacts of the model that one would not
expect to see in the Sun.
However, such structures were later also found in DNS \citep{BKKMR},
highlighting therefore the strong predictive power of MFS.

\begin{figure}[t!]\begin{center}
\includegraphics[width=\columnwidth]{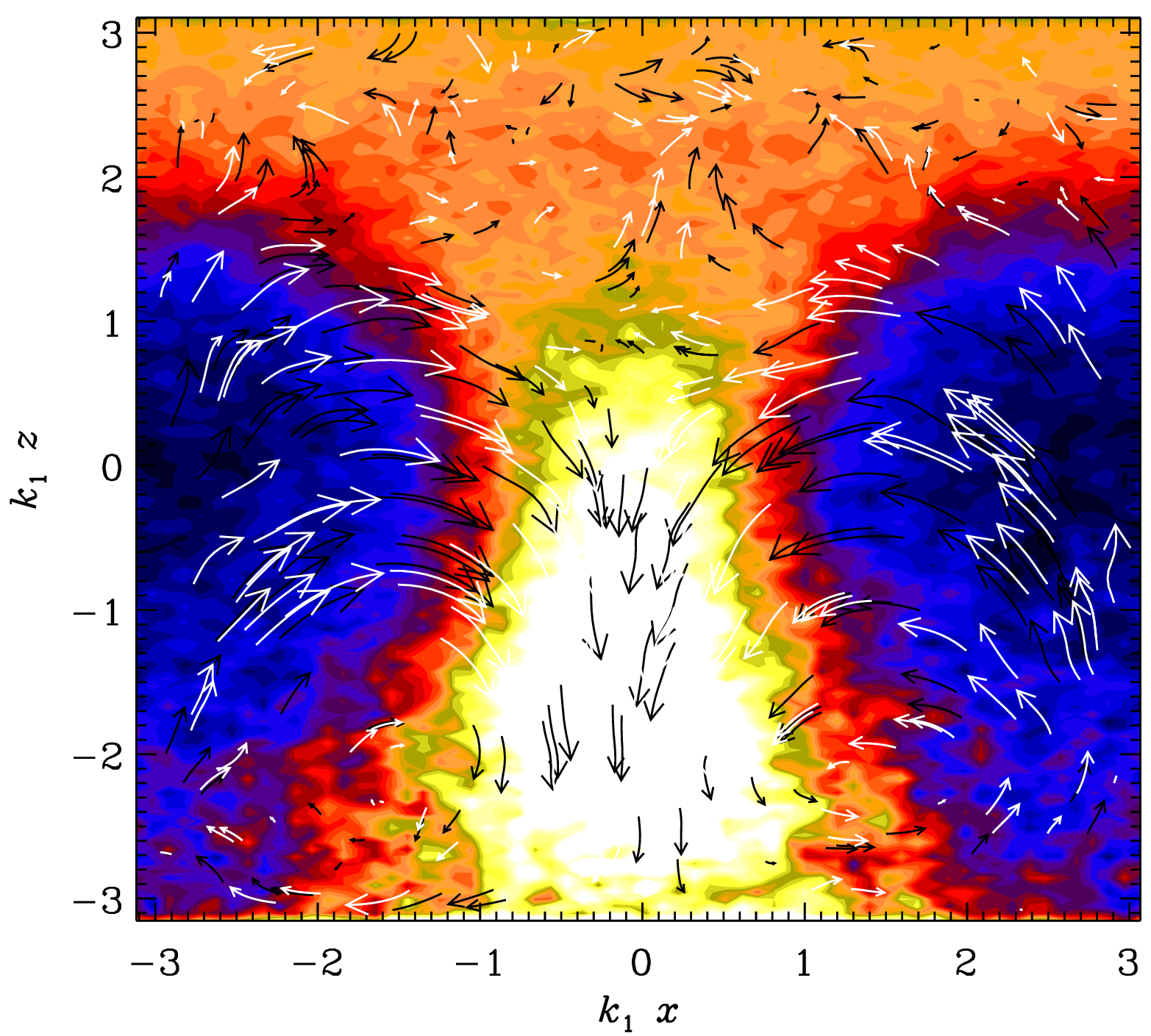}
\end{center}\caption[]{
Vectors of $\meanUU$ together with a color/grey scale representation of
$\meanB_y$ from DNS at a late time for $\Rm=18$ and $\Pm=0.5$.
}\label{DNS_U}\end{figure}

Visualizations of the resulting mean flow $\meanUU$ are shown in \Fig{DNS_U}
as vectors.
The flow shows a convergent shape toward the magnetic structures.
It is interesting to note that such convergent flow structures are now also
seen in local helioseismic flow measurements around active regions
\citep{Hindman}.
In this connection it is instructive to discuss the somewhat peculiar shape
of such a structure that widens as it descends.
Normally, in a strongly stratified atmosphere, descending structures
get compressed and become narrower, but this is not seen in the present
visualizations.
As already argued in \cite{BKKR12}, this is because the boundaries of these
structures do not coincide with material lines, so the mass is not
conserved inside them and can leak through the boundaries.
Indeed, these structures grow as they descend, and may become amenable
to helioseismic detection; cf.\ \cite{Ilonidis}.
This phenomenon is well known in the description of turbulent plumes
as a model of turbulent downdrafts in convection \citep{RZ95}.
Such structures are known to widen as a result of entrainment.
The sinking behavior of these apparently disconnected flow structures
can be explained as follows:
while inflows dominate downflows throughout the whole lifespan of the field concentration,
in the initial stage the former can drag in a large fraction of the surrounding magnetic field,
overcompensating the losses by downflows.
However as the environment gets depleted,
this dynamical balance shifts and the structures start moving downwards.

\subsection{Mean-field coefficients from DNS}

\begin{figure}[t!]\begin{center}
\includegraphics[width=\columnwidth]{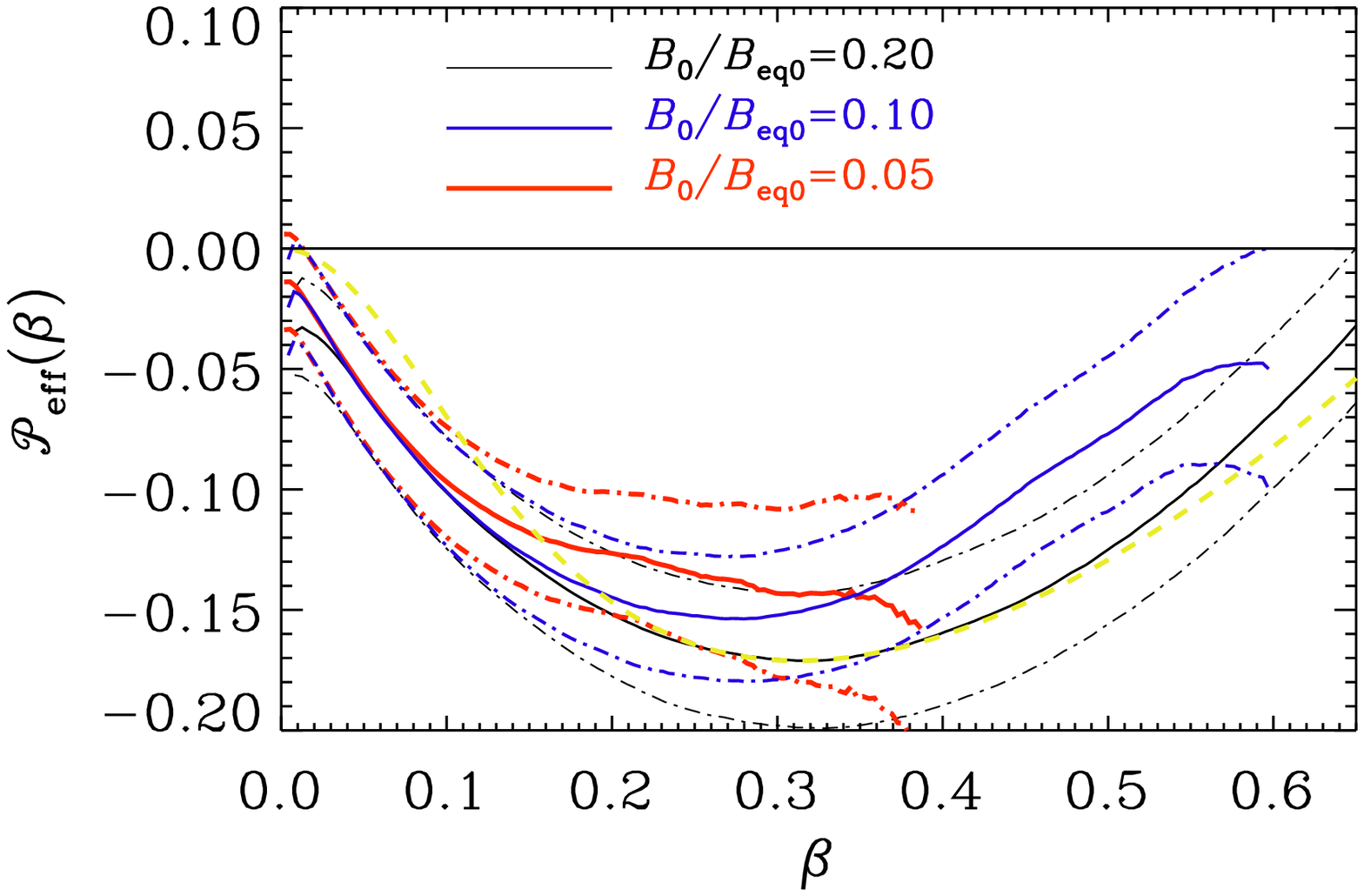}
\end{center}\caption[]{
$\Peff(\beta)$ for the DNS used in this paper with $\kf/k_1=30$,
$\Rm=18$, and $\Pm=0.5$.
}\label{DNS_qpc}\end{figure}

In earlier work by \cite{KBKR12}, the parameters $\qpz=40$ and
$\betap=0.05$, corresponding to $\betastar=0.32$, were used.
Those values are compatible with work by \cite{BKKR12} and \cite{KBKMR12}.
However, in the present case we have a larger scale separation ratio,
$\kf/k_1=30$, for which these parameters have not yet been determined.
In the \Fig{DNS_qpc} we show the functional form of $\Peff(\beta)$
for the present case with $\kf/k_1=30$, $\Rm=18$, and $\Pm=0.5$.
Here we have followed the method described in \cite{BKKR12};
see their Eq.~(17).
For the present model we find as fit parameters $\qpz=20$ and $\betap=0.167$,
which corresponds to $\betastar=0.75$.

\section{Comparison with MFS}

Recently, many aspects of NEMPI seen in the DNS have also been detected in MFS.
Establishing the usefulness and limitations of MFS is important, because
such models are easier to solve and allow one to explore parameters in
regimes where DNS are harder to apply or have not yet been applied in
the limited time span since the close correspondence between DNS and
MFS was first noted.

In the following we consider two-dimensional mean-field models, in which
the presence of $\qs$ has no effect on the solutions \citep{KBKR12}.
Furthermore, we ignore other effects connected with the anisotropy
of the turbulence.
These effects have previously been found to be weak \citep{BKKR12,KBKMR11}.

\subsection{Isothermal setup in MFS}
\label{modelMFS}

In this section we solve the evolution equations for mean velocity $\meanUU$,
mean density $\meanrho$, and mean vector potential $\meanAA$, in the form
\EQ
{\partial\meanUU\over\partial t}=-\meanUU\cdot\nab\meanUU
-\cs^2\nab\ln\meanrho+\grav+\meanFFFF_{\rm M}+\meanFFFF_{\rm K},
\EN
\EQ
{\partial\meanrho\over\partial t}=-\meanUU\cdot\nab\meanrho
-\meanrho\nab\cdot\meanUU,
\EN
\EQ
{\partial\meanAA\over\partial t}=\meanUU\times\meanBB-(\etat+\eta)\meanJJ,
\EN
where $\meanFFFF_{\rm M}$ is given by
\EQ
\meanrho \, \meanFFFF_{\rm M} = -\half\nab[(1-q_{\rm p})\meanBB^2]
+ \meanBB \cdot \nab\left[(1-q_{\rm s})\meanBB\right]\!,
\label{efforce}
\EN
and
\EQ
\meanFFFF_{\rm K}=(\nut+\nu)\left(\nabla^2\meanUU+\onethird\nab\nab\cdot\meanUU
+2\meanSSSS\nab\ln\meanrho\right)
\EN
is the total (turbulent plus microscopic) viscous force.
Here, $\meanSSS_{ij}=\half(\meanU_{i,j}+\meanU_{j,i})
-\onethird\delta_{ij}\nab\cdot\meanUU$
is the traceless rate of strain tensor of the mean flow.

We approximate $\qp$ by a simple profile that is only a function
of the ratio $\beta\equiv|\meanBB|/\Beq$.
We use an algebraic fit of the form
\begin{equation}
\qp(\beta)={\qpz\over1+\beta^2/\betap^2}.
\label{qpbeta2}
\end{equation}
The function $\qp$ quantifies the impact of the mean magnetic field
on the effective pressure force.

\subsection{Aspects of the MFS}
\label{MFSResults}

\begin{figure}[t!]\begin{center}
\includegraphics[width=\columnwidth]{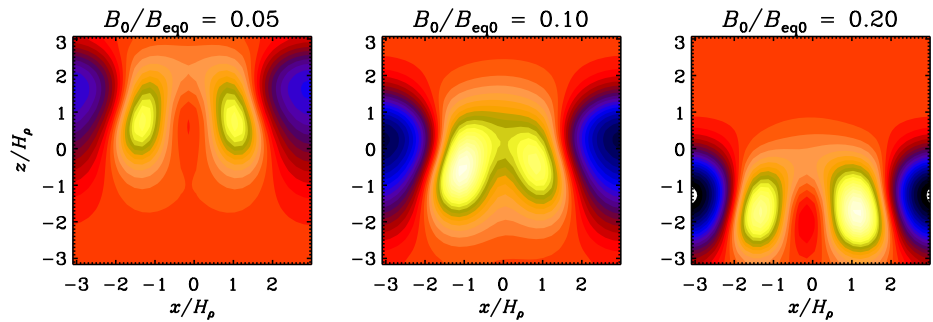}
\end{center}\caption[]{
$\meanB_y$ from mean-field models for three values of the imposed field strength
at the end of the linear growth phase of NEMPI.
Here, $\qpz=20$ and $\betap=0.167$, which corresponds to $\betastar=0.75$.
}\label{MF_B_stanb}\end{figure}

We begin by showing $\meanB_y$ for three values of the imposed field strength
at the end of the linear growth phase of NEMPI.
The results are shown in \Figs{MF_B_stanb}{MF_B} for two different setups.
In the former we use $\qpz=20$ and $\betap=0.167$ for the same $z$ range
($-\pi\leq z/H_\rho\leq\pi$) as in the DNS, while in the latter we use
$\qpz=40$ and $\betap=0.05$ for somewhat stronger fields and a deeper
$z$ range ($0\leq z/H_\rho\leq2\pi$), which is also the fiducial model used
by \cite{KBKR12}.
In the former case the growth rate is $\approx11 H_\rho^2/\eta_t$
while in the latter it is $\approx5.0 H_\rho^2/\eta_t$.

\begin{figure}[t!]\begin{center}
\includegraphics[width=\columnwidth]{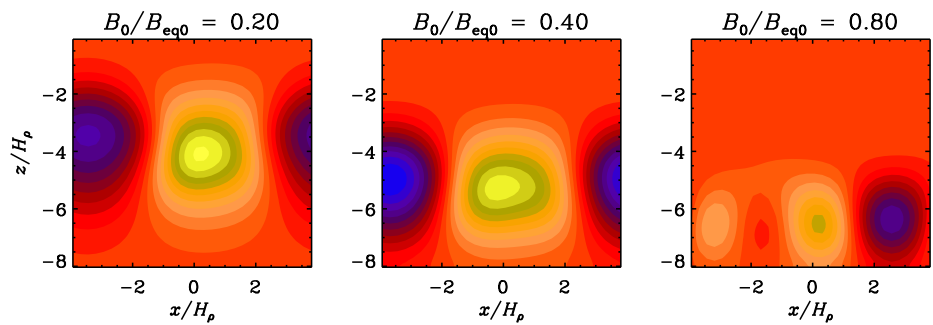}
\end{center}\caption[]{
Similar to \Fig{MF_B_stanb}, but for larger field strengths and vertical
domain boundaries that a deeper down, so the magnetic field maxima of the
instability fit better into the domain.
Here, $\qpz=40$ and $\betap=0.05$, corresponding to $\betastar=0.32$.
}\label{MF_B}\end{figure}

Unlike the DNS, the MFS show that in the former series of models with
$\qpz=20$ and $\betap=0.167$ the $x$ extend is slightly larger than the
optimal horizontal wavelength of the instability, because one sees
that some of the structures begin to split into two (\Fig{MF_B_stanb}).
This is not the case for the second model with $\qpz=40$ and $\betap=0.05$
(\Fig{MF_B}).

\begin{figure}[t!]\begin{center}
\includegraphics[width=.8\columnwidth]{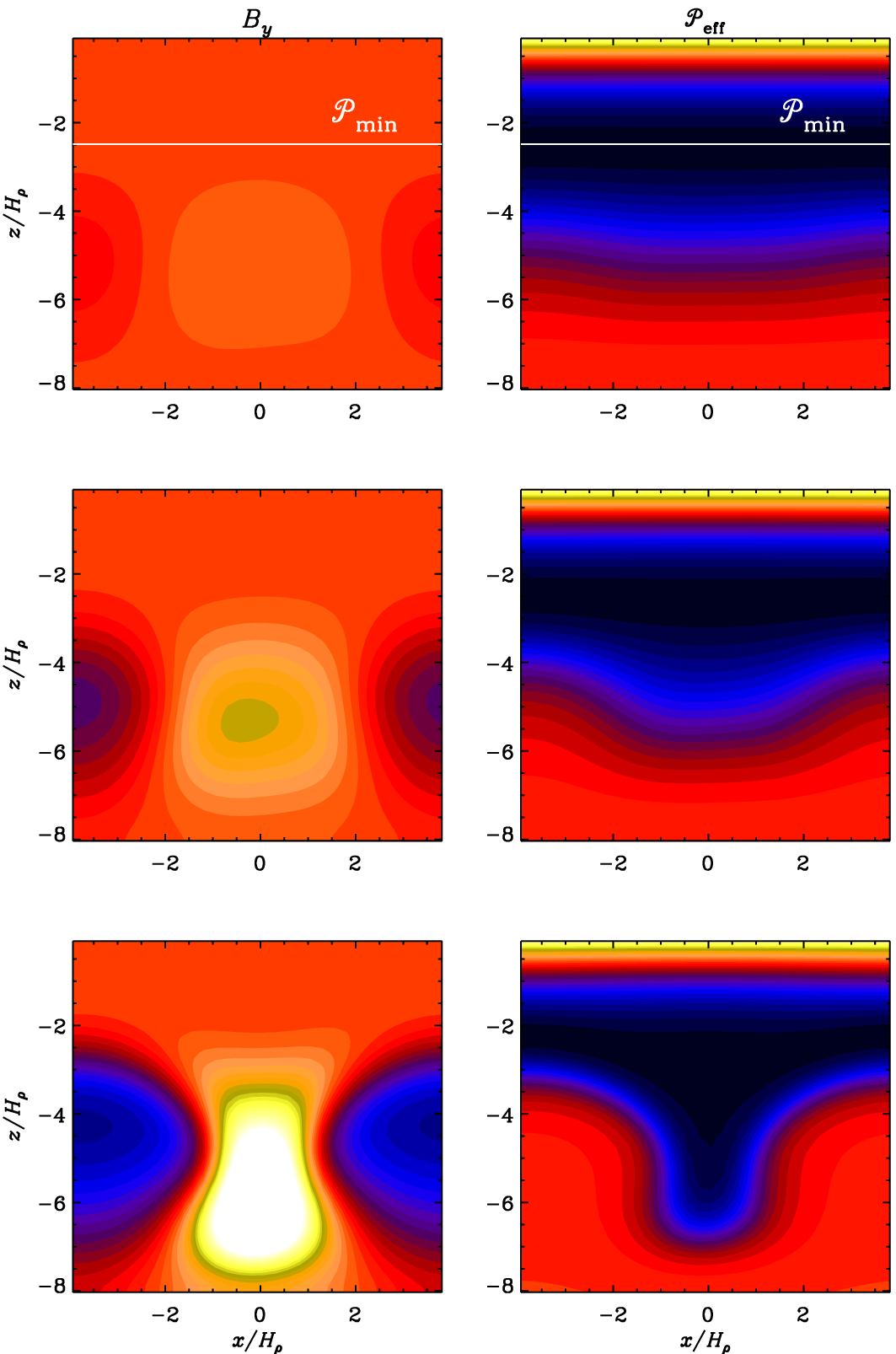}
\end{center}\caption[]{
$\meanB_y$ and $\Peff$ from mean-field models at three times.
}\label{MF_P}\end{figure}

\begin{figure}[t!]\begin{center}
\includegraphics[width=.8\columnwidth]{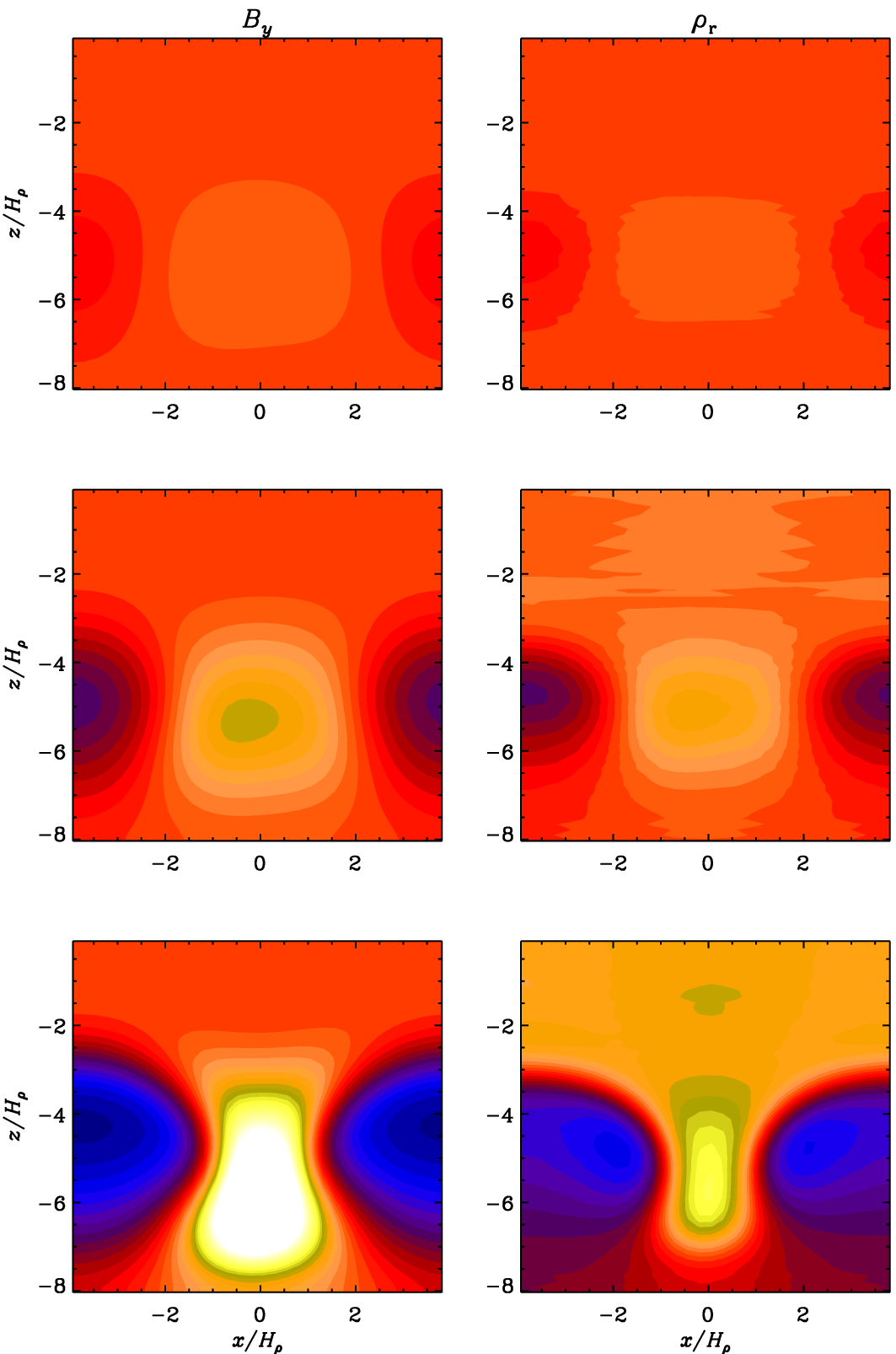}
\end{center}\caption[]{
$\meanB_y$ compared with $\meanrho$ for three different times.
}\label{MF_R}\end{figure}

Next, we compare $\meanB_y$ with $\Peff=\half(1-\qp)\beta^2$.
Again, there is a close correspondence between the $\meanB_y$ field
and the resulting distribution of $\Peff$; see \Fig{MF_P}.
Here, $\qp(\beta)$ is evaluated from the assumed fit formula given
by \Eq{qpbeta2}.
Furthermore, there is a close correspondence between regions of
enhanced magnetic field and enhanced density; see \Fig{MF_R}.

\section{Conclusions}

The present work has demonstrated that NEMPI is able to concentrate
the magnetic field into large patches encompassing the size of many
turbulent eddies.
The physics of this instability is a straightforward extension
of the usual magnetic buoyancy instabilities
\citep{P66,P79,HP88,Catt88,Wis00,Isobe05,Kers07}, except that the sign
of the buoyancy force is reversed in a regime of intermediate
field strength.
We have here re-examined the simple case of an isothermal layer
in which NEMPI can in principle occur at any depth whose value
is determined by the strength of the imposed field.
Our new DNS have verified that the growth rate is indeed independent
of the strength of the imposed field, provided the peak of the instability
fits still comfortably within the domain.
During the subsequent nonlinear evolution of the instability, the
overall density stratification readjusts, allowing the magnetic field
concentrations to move further down.
It is important to realize that the resulting structures are subject
to significant turbulent entrainment \citep{RZ95},
so their boundaries are not closed.
The agreement with corresponding mean field models is astounding
and much more convincing than what has so been possible to demonstrate
in mean-field dynamo theory.
Mean-field models provide therefore a strong source of guidance
when designing new setups for DNS.

While NEMPI now begins to be fairly well understood for isothermal models,
more work is required for non-isothermal ones.
In that case, the density scale height is no longer constant and the
degree of stratification is much stronger at the top than in deeper layers.
The simple result that the instability can occur at any height, depending
just on the strength of the imposed field, is then no longer valid.
At the same time, there is another perhaps more important aspect,
namely the possibility of other instabilities.
One of them is connected with the suppression of turbulent convective
energy flux by the mean magnetic field.
As shown by \cite{KM00}, this effect can also lead to magnetic flux
concentrations and it may be sustained for much stronger magnetic
field strengths, allowing thus the formation of structures in which the
magnetic pressure becomes comparable to the ambient gas pressure.
There may also be a connection with flux segregation events seen in
simulations of magnetoconvection at large aspect ratios \citep{Tao,KKWM10},
which have already been shown to produce bipolar regions in simulations
with radiation transfer \citep{Stein}.
The study of the possibility of producing sunspots similar to those
of \cite{Rem11a,Rem11b}, but without initial flux structures,
is now of high priority in the quest for
solving the solar dynamo problem in terms of distributed dynamo models
in which magnetic activity is explained as a surface phenomenon.

\begin{acks}
Computing resources provided by the Swedish National Allocations Committee
at the Center for Parallel Computers at the Royal Institute of Technology in
Stockholm and the High Performance Computing Center North in Ume{\aa}.
This work was supported in part by the European Research Council
under the AstroDyn Research Project No.\ 227952
and the Swedish Research Council under the project grant 621-2011-5076.
\end{acks}

\appendix

\section{Growth rate of NEMPI}
\label{LinTheo}

In this Appendix we derive the growth rate of NEMPI neglecting for simplicity
dissipation processes,
using anelastic approximation for small Mach numbers
and assuming the density hight, $H_\rho$ to be constant and $\mu_0=1$.
Let us rewrite the equation of motion in the following form
\EQ
{\partial\meanUU(t,x,z)\over\partial t}=-\nab \left( {p_{\rm tot} \over \meanrho}\right)
+ {p_{\rm tot} \over \meanrho H_\rho} \hat{\bm z}+\grav,
\label{A1}
\EN
where $\hat{\bm z}$ is the vertical unit vector,
$p_{\rm tot}=\meanp + p_{\rm eff}$ is the total
pressure (the sum of the mean gas pressure, $\meanp$, and the effective magnetic pressure,
$p_{\rm eff}$), and we took into account that mean magnetic field is independent on $y$,
so that the mean magnetic tension vanishes.
We also used an identity:
\EQ
\nab\left({p_{\rm tot} \over \meanrho}\right) = {\nab p_{\rm tot} \over \meanrho}
+ {p_{\rm tot} \over \meanrho H_\rho} \hat{\bm z},
\label{A4}
\EN
Taking twice curl of \Eq{A1} we obtain
\EQ
{\partial\over\partial t} \left(\Delta - {\nabla_z \over H_\rho}\right) \meanU_z=
 {1 \over \meanrho H_\rho} \Delta_\perp p_{\rm eff},
\label{A2}
\EN
where $\Delta_\perp= \Delta-\nabla_z^2$ and we have used \Eq{divu}.
Introducing a new variable $V_z=\sqrt{\meanrho}\,\meanU_z$,
we rewrite \Eq{A2} for a new variable:
\EQ
{\partial\over\partial t} \left(\Delta - {1 \over 4 H_\rho^2}\right) V_z=
 {1 \over \sqrt{\meanrho} H_\rho} \Delta_\perp p_{\rm eff} .
\label{A5}
\EN
Linearizing \Eq{A5}, and using the linearized induction
\Eq{induct-eq} we arrive at the following equation:
\EQ
{\partial^2\over\partial t^2} \left(\Delta - {1 \over 4 H_\rho^2}\right) V_z(t,x,z)=
- {2 \beta_0^2 \over \meanrho H_\rho^2}
\left.{\dd p_{\rm eff}\over\dd\beta^2}\right|_{\beta_0} \Delta_\perp V_z,
\label{A6}
\EN
where $\dd p_{\rm eff}/\dd\beta^2=\half(1-\qp-\dd\qp/\dd\ln\beta^2)
B_0^2/\beta^2$, $\beta=\meanB/\Beq(z)$ and $\beta_0=B_0/\Beqz$.
It follows from \Eq{A6} that a
necessary condition for the large-scale instability is
\EQ
\left.{\dd p_{\rm eff}\over\dd\beta^2}\right|_{\beta_0} < 0 .
\label{A10}
\EN
For instance, in WKB approximation when
$k_z \, H_\rho \gg 1$, i.e. when the characteristic scale of the spatial variation
of the perturbations of the magnetic and velocity fields are much smaller than the
density hight length $H_\rho$, the growth rate of the instability reads
\EQ
\lambda= {\beta_0 k_\perp \over \sqrt{\meanrho_0} \, H_\rho k} \,
\left(-2 \left.{\dd p_{\rm eff}\over\dd\beta^2}\right|_{\beta_0} \right)^{1/2} .
\label{A11}
\EN
For an arbitrary $k_z \, H_\rho$ we seek a solution of \Eq{A6}
in the form: $V_z(t,x,z) = V(z) \exp(\lambda t+ i k_\perp \,x)$.
Introducing new variables:
\EQ
\Psi(R)=\sqrt{R} \, V(z), \quad
R(z) ={v_{A0}^2 \over u_{\rm rms}^2 \betap^2} \,e^{z/H_\rho} ,
\label{A7}
\EN
we can rewrite \Eq{A6} in the form of a 1-D Schr\"odinger equation
for the function $\Psi(R)$:
\EQ
{d^2 \Psi\over dR^2} - \tilde U(R) \Psi =0, \quad
\tilde U(R)= {k_\perp^2 \over R} \, \left({H_\rho^2\over R}
- {a \, \qpz\over (1+R)^2} + a\right),
\label{A8}
\EN
where $\meanrho=\meanrho_0 \, e^{-z/H_\rho}$, $v_{\rm A0}=B_0/\sqrt{\meanrho_0}$
is the Alfv\'en speed, the parameter $a$ is
\EQ
a={u_{\rm rms}^2 \, \betap^2 \over \lambda^2} ,
\label{A9}
\EN
and the potential $\tilde U(R)$ has the following asymptotic behavior:
$\tilde U(R\to 0)=k_\perp^2 H_\rho^2/ R^2$ and
$\tilde U(R\to \infty)=a/R$.
For the existing of the instability, the potential $\tilde U(R)$ should
have a negative minimum.
For example, for a long wavelength instability ($k_\perp^2 H_\rho^2 \ll 1$) and
when $\qpz>1$, the potential $\tilde U(R)$ has a negative minimum,
and the instability can be excited.
When the potential $\tilde U(R)$ has a negative minimum and since
$\tilde U(R\to 0)>0$ and
$\tilde U(R\to \infty)>0$, there are two points
$R_1$ and $R_2$ (the so-called turning points) in which $\tilde U(R)=0$.
Using the equations $\tilde U(R_{1,2})=0$ and Equations~(\ref{A8})--(\ref{A9})
we obtain the growth rate of the instability as
\EQ
\lambda= {\betastar \,u_{\rm rms} \over H_\rho} \,
{\left[R_1 R_2 (2+R_1+R_2) \right]^{1/2}\over (1+R_1)(1+R_2)} ,
\label{A12}
\EN
where we have used $\betastar=\betap \sqrt{\qpz}$. Note that \Eq{A12}
is consistent with the simple estimate~(\ref{E1}).

\end{article}
\end{document}